# Multipacting Simulations of Tuner-adjustable waveguide coupler (TaCo) with CST

*Nuaman Shafqat \*, Frank Gerigk, Rolf Wegner (CERN)*
*Pakistan Atomic Energy Commission, Islamabad, Pakistan

**Keywords:** Multipacting, Integrated Secondary yield, coupler, waveguide coupler, adjustable coupler

**Abstract**

Tuner-adjustable waveguide couplers (TaCo) are used to feed microwave power to different RF structures of LINAC4. This paper studies the multipacting phenomenon for TaCo using the PIC solver of CST PS. Simulations are performed for complete field sweeps and results are analysed.

# Contents

# 1 Introduction

In RF equipment with electric surface fields electrons can be emitted from the surface and accelerated. On surface impact secondary electrons can be produced, which are again accelerated in the RF field. At certain field levels it can happen that the particle movement and the RF are in synchronism and if the Secondary (Electron) Emission Yield (SEY) is larger than 1 an exponential multiplication of electrons [1] can take place, called multipacting. Multipacting is often found at lower field levels and after certain time of RF conditioning the SEY goes down and these multipacting barriers can be surpassed. However, if multipacting occurs at field levels close to normal operating fields, then the operation may be severely impacted and even damage of structure is possible.

A tuner-adjustable waveguide coupler (TaCo), which is a modification of a $\lambda/4$ waveguide coupler, is used for coupling RF power from the klystron to the RF structures of LINAC4. TaCo is equipped with single slug tuner to adjust the cavity-to-waveguide coupling [2]. The iris has a racetrack with 50mm width making it vulnerable for multipacting.

Different analytical techniques are used to study multipacting behaviour but to a great extent these techniques are based on a one- dimensional model with spatially uniform electromagnetic fields. On the other hand most of RF devices involve structures with inhomogeneous electromagnetic fields, analytical techniques can only give very rough estimates. This makes numerical techniques essential for reliable and accurate prediction of multipacting barriers in RF structures [3].

CST Particle Studio is a powerful tool to study multipacting in RF structures. It allows to import fields calculated with the CST Eigen-mode Solver and to track the particles inside the RF structure. It is equipped with different probabilistic emission models and gives flexibility to modify emission properties as well. In the following CST Particle In Cell (PIC) is used for multipacting study of the iris of TaCo

# 2 Multipacting

Multipacting (MP) is a resonant RF electron discharge in which electron multiplication takes place due to the secondary electron re-emission process. The discharge is mainly encountered in RF accelerating structures where the combination of RF fields and clean surfaces of high secondary yield metals like copper, aluminium, or niobium will increase the number of electrons at each impact

## 2.1 Basic Theory

For multipacting to take place the following two conditions are necessary

1. Synchronization of electron movement with the RF fields
2. Electron multiplication via secondary electron emission

The first condition is met when the time between two electron impacts is an integer multiple of half of an RF period. In this case the emitted electrons will have exactly same starting conditions resulting in sustained oscillations of an electron cloud which will keep on growing as governed by second condition.

The second condition defines the mechanism of how the electrons will grow upon each impact.

For multipacting to take place the secondary emission coefficient δ has to be greater than one. Another parameter that governs the multipacting process is impact energy U. When a primary electron collides with a surface then it depends on its impact energy whether it is:

- Reflected
- Absorbed, or
- cause the emission of secondary electrons.

For secondary electron multipaction to occur the impact energy U of primary electrons for metals is usually in the range of 50 eV to 1500 eV.

## 2.2 Types of Multipacting

Multipacting can be broadly divided into three different types

- Two-point multipacting in short-gap cavities $L \ll \lambda$
- One-point multipacting at the cavity equator of axial-symmetric cavities
- Two-point multipacting at the cavity equator of axial-symmetric cavities

### 2.2.1 *Two-point multipacting in a short-gap $L \ll \lambda$*

This type of multipacting occurs in short-gap cavities when the time of flight of electron is an odd integer multiple of half of RF period.

Two-point multipacting can be explained using the example of a field in a parallel-plate capacitor with gap length L. Here the multipacting barriers can be predicted using the following expressions [1].

$$V = L.E_o = 4\pi \frac{m_0 c^2}{e} \left(\frac{L}{\lambda}\right)^2 \frac{1}{(2n-1)}$$

$$U = 8m_o c^2 \left(\frac{L}{\lambda}\right)^2 \left(\frac{1}{2n-1}\right)^2$$

Where L is the gap length and n is order of multipacting barrier.

### 2.2.2 *One-point multipacting at the cavity equator for axial-symmetric cavities*

This type of discharge occurs when electrons impact the cavity surface more or less at the same place. For multipacting to take place the time of flight of electrons must be around an integer multiple of the RF period so that emitted electrons will have exactly the same set of initial conditions. This type of discharge often takes place in areas where the magnetic field is strong enough to govern the multipacting process.

For this type of multipacting there is no simplified analytical model available and the energy gain around the trajectory is strongly dependent on the details of the field distribution. However [1] suggests to estimate multipacting barriers by using following approximate formula.

$$B_o = \frac{2\pi \nu m_o}{e} \frac{1}{n}$$

Where $\nu$ is RF frequency and n is the order of the multipacting barrier.

### 2.2.3 *Two-point multipacting at the cavity equator of axial-symmetric cavities*

This type of multipacting occurs when secondary electrons emitted by the cavity surface near equator are bent in cyclotron-like trajectory and fly across the cavity mid-plane and strike the cavity surface at a point nearly symmetric to starting point. This phenomenon was first observed at CERN in prototype cavities for LEP project.

No simplified analytical model exists for such type of multipacting but the formula from above can be modified for a rough prediction of the multipacting barriers [1].

$$B_o = 2\frac{2\pi\nu m_0}{e}\frac{1}{(2n-1)}$$

Where $\nu$ is RF frequency and n is multipacting order.

## 3 Tuner-adjustable waveguide coupler (TaCo)

Tuner-adjustable waveguide coupler (TaCo) in Figure 1 is a modification of the T-type $\lambda/4$ waveguide coupler with fixed short circuit plate and a slug tuner to vary the cavity-to-waveguide coupling.

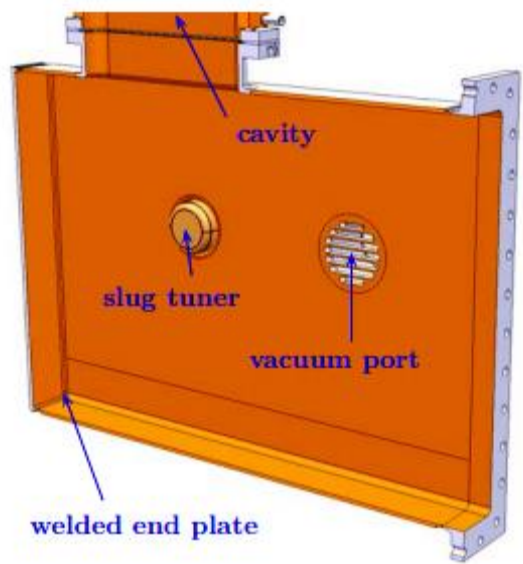

**Figure 1: Tuner-adjustable waveguide coupler**

### 3.1 Salient Features of TaCo

TaCo has significant advantages when compared with conventional T-type couplers [2]

- Short circuit plate is fixed which considerably facilitates machining process and saves two waveguide flanges and one gasket.
- Coupling depends on magnetic field strength in iris towards cavity.
- In a regular T-type coupler the cavity-to-waveguide coupling can be modified by replacing the short-circuit while in TaCo the waveguide-to-cavity coupling can be adjusted by changing the tuner penetration, making the tuning process very convenient.

## 4 Multipacting Setup for Simulation

Different numerical simulation codes are used for predicting multipacting. Every code has certain set of advantages coupled with certain limitations. The most common problem is that most codes are 1D and 2D and use a semi-empirical approach derived by Vaughan [4] where electron reflection is not accounted. Some other codes [5] [6] which incorporate electron reflection suffer due to very limited applications for certain vacuum electronic devices like photomultiplier tubes, collectors in liner beam tubes and distributed emission crossed-field amplifiers (CFA's). Also these codes use macro-particles instead of individual electrons, where a large number of electrons are located at the same point, with identical momentum. The use of macro-particles means that information about the energy distribution cannot accurately resolved and statistical nature of electron reflection and emission is at the best integrated over and smoothed out. For design and analysis of RF structures with complex geometries and non-uniform electromagnetic fields it is required to use a software tool, which combines electromagnetic field simulation, multi-particle tracking and advanced probabilistic emission models.

## 4.1 CST Particle Studio (PS)

CST particle studio has following advantages for realistic multipacting simulations

- Electromagnetic field calculation using the CST Eigen-mode Solver the results of which can be directly imported into CST PS.

- Multi-particle tracking using Particle–In-Cell (PIC) or Particle Tracking solver (TRK).

- Incorporates advanced emission model, which was introduced recently by Furman and Pivi [7] and which also includes electron reflections.

- Incorporates the flexibility to accommodate user defined inputs even for the secondary emission model.

- Vast post processing options to analyse the results.

All these features make CST PS a very attractive tool for the accurate study of multipacting in complex RF structures with inhomogeneous electromagnetic fields.

## 4.2 Problem definition

In order to study and analyse the multipacting behaviour of TaCo we start with the field simulation of the coupler and then continue with the multipacting simulation and analysis.

## 4.3 Field Simulation of Tuner-adjustable waveguide Coupler (TaCo)

The Eigen-mode solver of CST with very dense rectangular mesh of 70 LPW was used for calculation of the fields in the tuner-adjustable waveguide coupler. To simplify the calculation the actual Linac4 cavity was replaced with a dummy cavity having the following characteristics:

- Operating frequency should be around 352.2MHz.

- The coupling should be in the range of 10 to 15 (to simulate the coupling to an unloaded 7-cell cavity).

- The input port must be modelled as perfectly matched load with no reflection.

### *4.3.1 Eigen Mode Setup for Field Simulation*

#### *4.3.1.1 Mesh Definition*

In order to limit the simulation time a certain effort was made to optimise the meshing of problem space.

A rectangular mesh of very high density in the area of interest was used for the field calculation in coupler. Reason to use rectangular mesh stems from the fact that CST PS only supports rectangular mesh. Very dense mesh requirement is needed as we want to have very accurate definition of the fields especially near the surface as this is required for reliable and accurate multipacting simulations. Figure 2 shows the increased mesh density in the area of coupling slot.

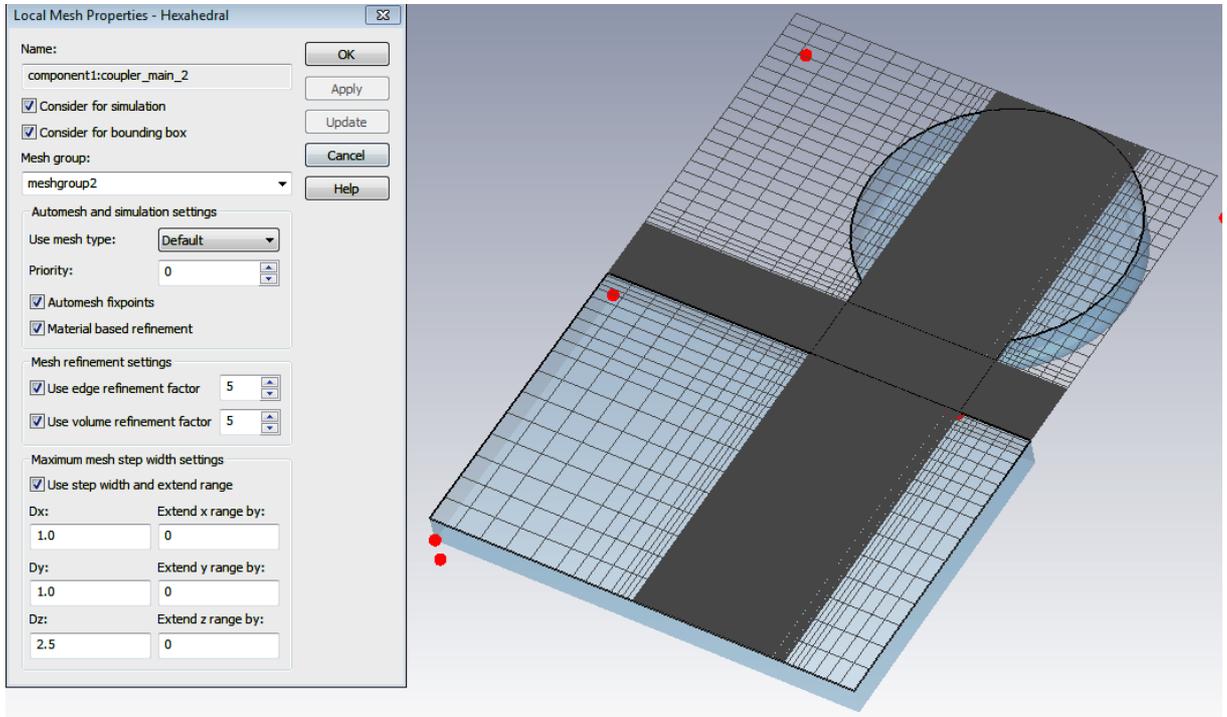

**Figure 2: Mesh Definition with Dense Mesh in Area of Interest**

*4.3.1.2   Modelling of Perfect Absorber in CST*

Ports in CST can be defined as perfectly matched loads with zero reflection so this was the choice for the input port (right hand side of Figure 1).

*4.3.1.3   Adjustment of Coupling and Frequency*

CST allows to define certain parameters of the structure as variables, which was used to adjust different parameters like coupling, Q value and frequency.

*4.3.1.4   Calculation of the Coupling*

CST allows calculating Q and external Q with its post processing tools:

$$Coupling = \frac{Q}{Q_{ext}}$$

External Q was adjusted by changing parameters of the cavity while cavity Q was adjusted by variation of surface resistance.

*4.3.1.5   Simulation Results*

The electric field pattern in TaCo simulated with CST is shown in Figure 3, while different simulation results are summarised in Table 1.

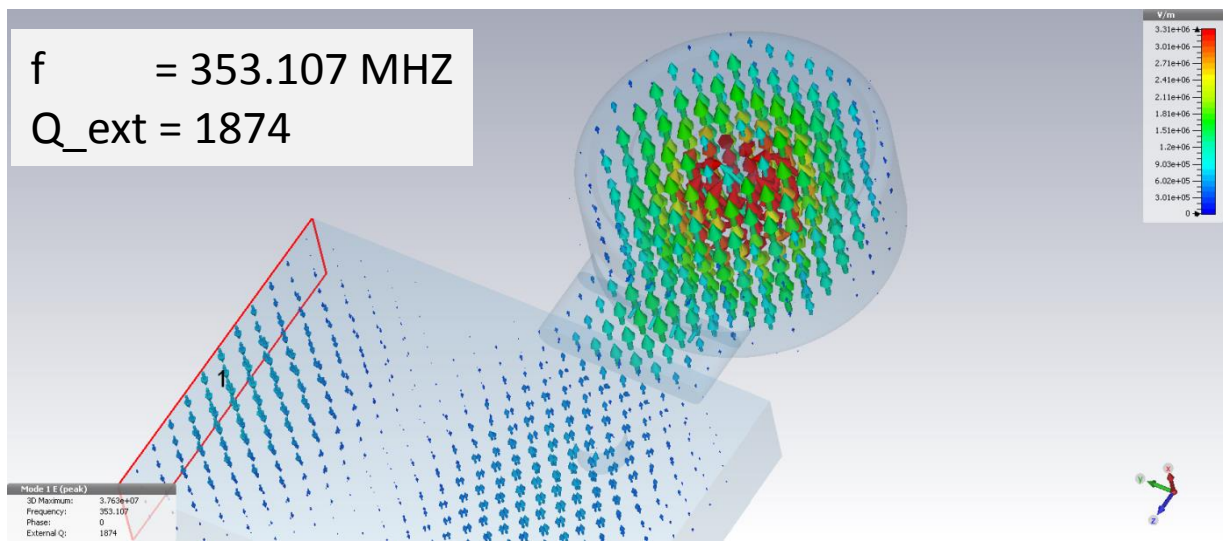

**Figure 3: Electric Field Plot**

**Table 1:**

| Simulation Results | |
|---|---|
| Frequency | 353.107MHz |
| Q | 2607 |
| Q_ext | 1874 |

## 4.4 Multipacting Simulations of Tuner-adjustable Waveguide Coupler (TaCo)

### 4.4.1 Particle Studio Setup for Multipacting Simulations

Particle Studio of CST is equipped with two solvers one is Particle-In-Cell (PIC) and other is Particle Tracking (TRK) both can be used for multi particle tracking. I used PIC for Multipacting Simulation due to following reasons.

– More Multipacting Specific infrastructure

– Better memory handling than in TRK

– Space charge effect can be ignored

#### 4.4.1.1 Modification of Structure

The Eigen-mode solver takes a Perfect Electric Conductor (PEC) as boundary conditions and does not support any type of lossy material. So for the multipacting simulations there are two options: i) either to rebuild the structure from scratch or ii) to import the model from Eigen-mode solver and shell it with the material of choice from the CST material library. In the following I used the second approach.

#### 4.4.1.2 Secondary Electron Emission Model

The CST Material Library contains a wide choice of materials with pre-defined secondary emission properties based on the Furman Model. Different secondary emission properties of materials can be changed if needed. I used pre-defined SEE-Copper for simulations.

*4.4.1.3 Definition of Particle Source*

CST PS allows to define multiple types of particle sources. The most important ones are: i)the "Particle Point Source", which allows to define particle sources at any point in the structure., ii)"Particle Source on any Surface", which allows to define multiple particle point sources on any critical surface of our choice.

Different types of emission models are available in PIC like Gaussian, DC, Field and Explosive but Gaussian is best choice for multipacting simulations as it produces user controlled bunches of particles which can be switched off after a certain time. For multipacting simulations only an initial seed is required after which process follows the dynamics of the fields and the secondary emission properties of the chosen material.

Particle sources were defined in the iris of TaCo with twenty eight emission points and a Gaussian Emission was chosen with an initial energy of 5 eV as shown in Figure 4.

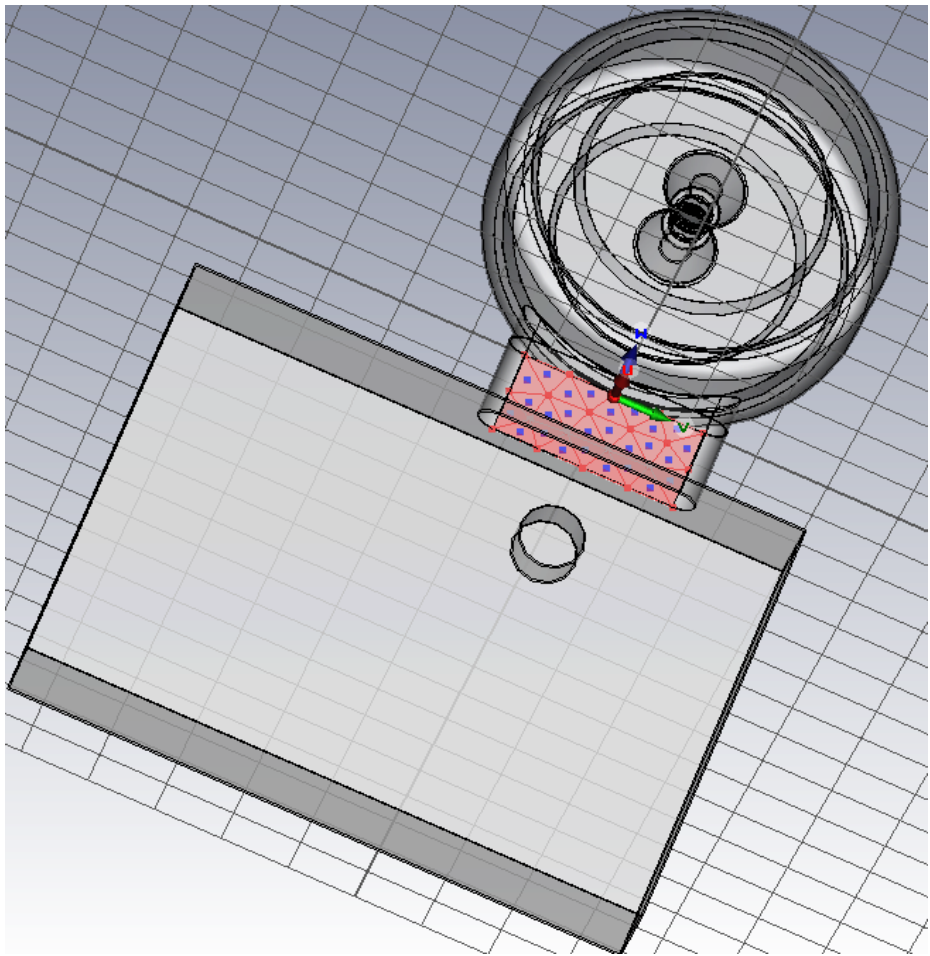

**Figure 4:Definition of Particle Source in Iris of TaCo**

*4.4.1.4 How to Apply Input Power*

CST PS allows using two different schemes for powering the structure. One is to apply the input power via the input port but this has the disadvantages of time delay and reaching steady state. The second and the most recommended way is to import the fields calculated with the Eigen Solver, which can be scaled to the required field values.

For these simulations the electric and magnetic fields were imported from Eigen Solver.

*4.4.1.5  Scaling of Imported Field*

Eigen-mode fields are normalized to one Joule instead of one watt so scaling has to be adjusted accordingly. In order to calculate the scaling factor of the electric field on axis of the dummy cavity is determined and compared with the actual field gradient of real Linac4 cavities to calculate the scaling factor. This operation is done with post processing tools.

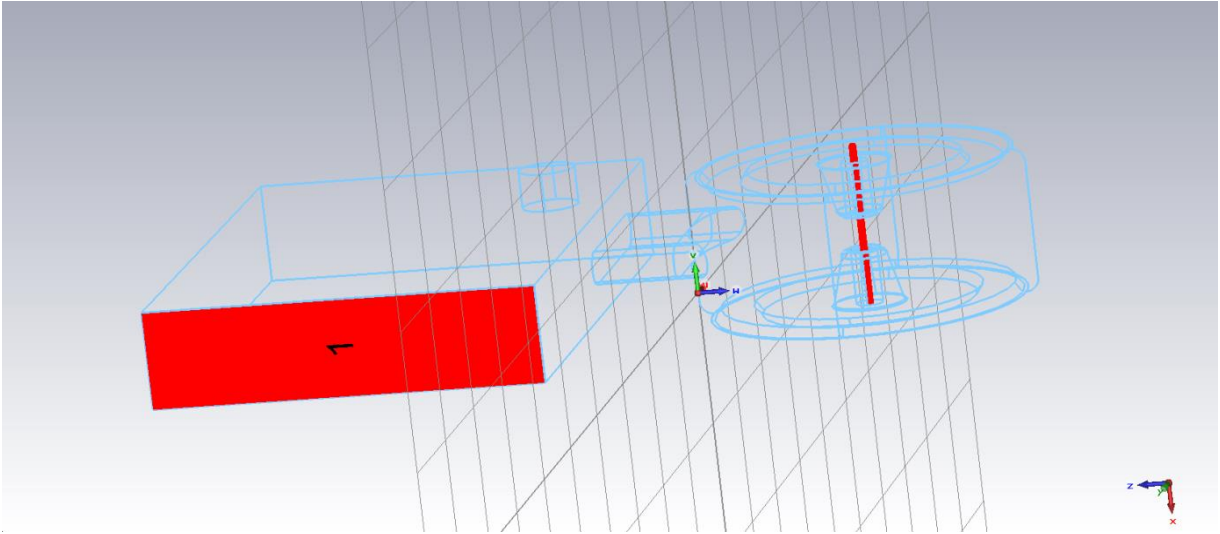

**Figure 5: Setup for calculation of the on axis field**

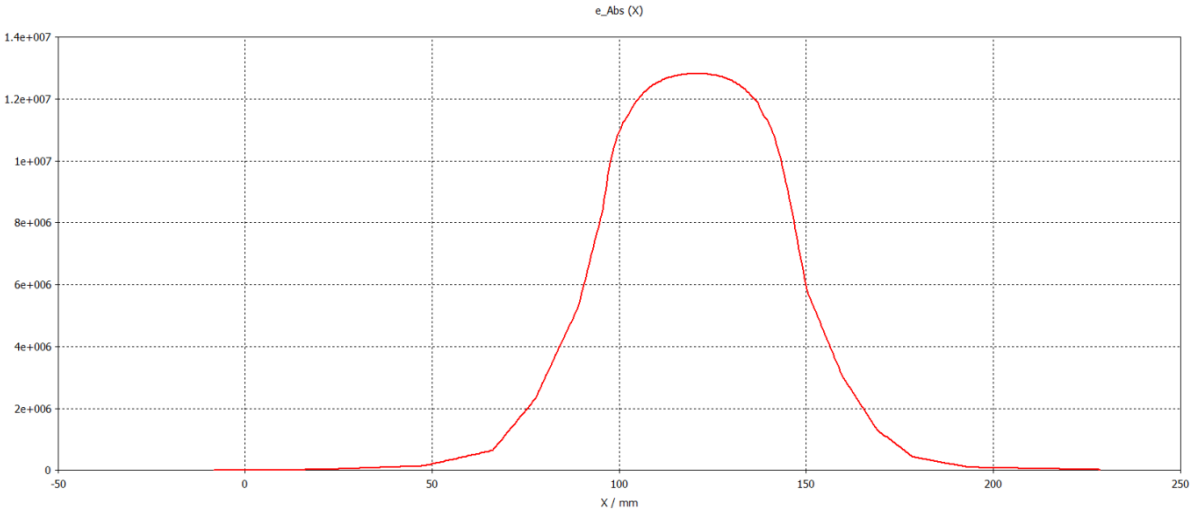

**Figure 6: On axis field profile**

Now by integrating the field in Figure 5 and comparing it with nominal Linac4 PIMS gradient of 4Mv/m one gets a scaling factor of 1.156 which has to be applied to externally imported fields.

*4.4.1.6    Mesh Definition*

CST PS uses the first mesh layer above the surface to emit the electron. At high retarding electric fields this layer may be large enough for electrons to gain energy large enough to produce true secondaries resulting into rapid unphysical growth of electrons. This phenomenon occurs mostly at and near the surface if the mesh size is bigger than the average length of the electron trajectories and if the filed there is not smooth enough. The stronger the filed, the shorter the trajectories become requiring more and more dense mesh. An extensive discussion about the mesh size is made in [3] and [9]- as very dense mesh require very powerful computation resources and very long simulation times.

As a rule of thumb an unphysical rapid growth means that mesh density must be increased and for higher fields it become mandatory to choose denser meshes in the area of interests for both field calculation in Eigen Solver and multipacting simulations in PIC Solver.

*4.4.2    Multipacting Tools in PIC*

To detect multipacting is not a trivial process as multipacting is not established until a sizeable electron current is produced which may require large number of RF cycles.

The following post processing tools are available with PIC of PS.

*4.4.2.1    Particle vs Time*

Particle vs Time curve shows how the particles are increasing with time. This diagnostic is automatically given as post processing plot at the end of simulation.

*4.4.2.2    Emitted Secondaries*

This is very important tool to analyse type of multipacting process and its trend as it shows the hits and resultant emitted secondaries. This diagnostic tool is automatically given at the end of simulation.

*4.4.2.3    Collision and Emission Information*

Automatically available as post processing result which contains the information about collision current, collision power, emission current and emission power for each component of structure. This information is needed to calculate the average or integrated secondary yield defining the chances of multipacting at different field levels of interest.

*4.4.2.4    PIC Position Monitor*

One of the most visual tool of the PS PIC solver, which allows you to see how the particles are moving in space and their collision and resultant emitted particles. PIC Position-Monitor is not automatically available post processing result and we have to add it.

*4.4.2.5    PIC Phase Space Monitor*

The secondary emission, which defines the multipacting process, is highly dependent on the energy electrons gain before collision. With this tool we can see the energy gain of the electrons and their movement in any co-ordinate axis at any given time. PIC-Phase-Monitor is not automatically available post processing result and we have to add it.

#### 4.4.2.6 Integrated or Average Secondary Yield (<SEY>)

The integrated or average secondary yield defines how the number of the electrons increase per particle crossing [8]. A plot of <SEY> vs field strength clearly defines the critical field levels and the areas in structure susceptible for multipacting. If <SEY> is greater than one it means that the number of particles will grow with time.

<SEY> is not directly available in the PIC post processing tool but it can be very easily calculated from the collision and emission current information.

#### 4.4.2.7 Few Important Considerations

Multipacting currents build up gradually over many RF cycles so a sufficient number of collisions must be considered for accurate results.

### 4.4.3 Multipacting Results and Analysis

For the operating frequency of 352.2 MHz it was decided to use 100 ns simulation time (35 RF cycles) to make sure that a sufficient number of collisions is computed.

#### 4.4.3.1 Integrated or Average Secondary Electron Yield <SEY> Curve

Figure 7 gives plot of <SEY> vs field with 60 data points for a full field sweep.

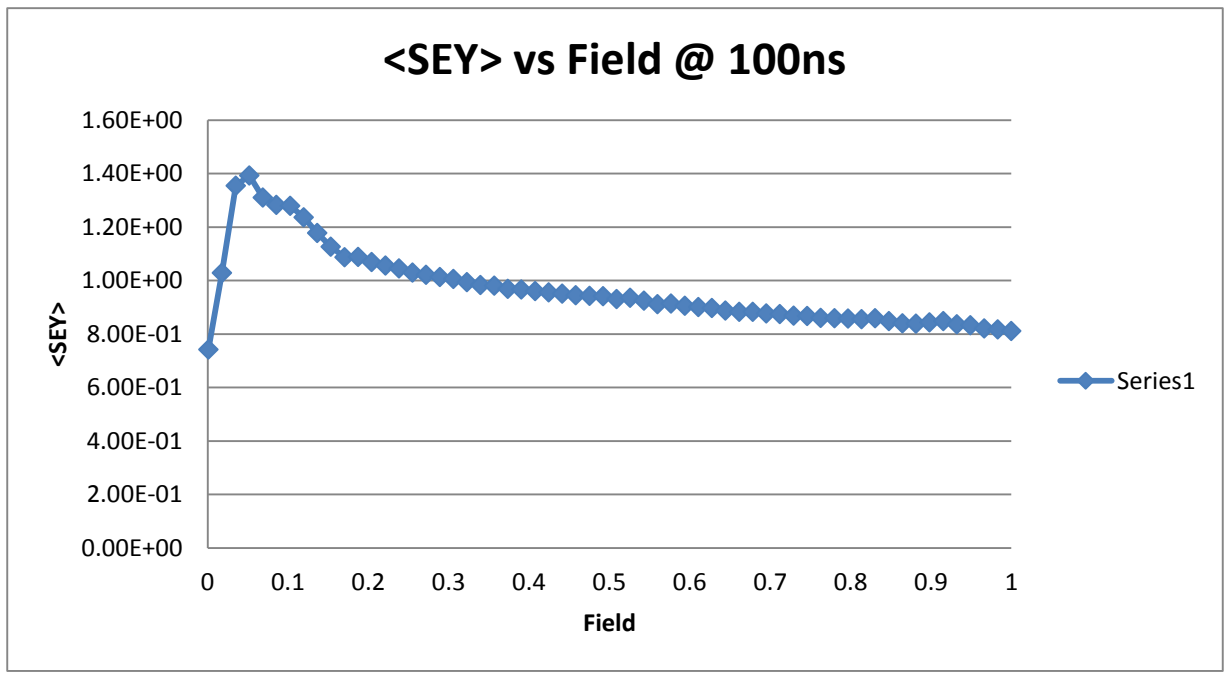

**Figure 7: <SEY> vs Field**

#### 4.4.3.2 Explanation of <SEY> Curve

Using equations for Two Point Multipacting we can calculate the two point multipacting barriers in the iris of TaCo with 50 mm width.

**Table 2: First five Two Point Multipacting Barrier in the iris of TaCo (analytical estimate)**

| Barrier Order n | E [V/m] |
|---|---|
| 1st | 445.50E+3 |
| 2nd | 148.50E+3 |
| 3rd | 89.10E+3 |
| 4th | 63.60E+3 |
| 5th | 49.50E+3 |

*4.4.3.3  Electric Filed in Iris of TaCo*

From Figure 8 it is clear that the electric field in the iris with (calculated) maximum of 2.23MV/m is not a uniform capacitive field as assumed for the derivation of the formulas for the prediction of two point multipacting barriers, so we can expect some deviation from the calculated multipacting barriers. We can expect that first and second order multipacting barrier should occur around 0.1996 and 0.0665 times the filed maxima in slot.

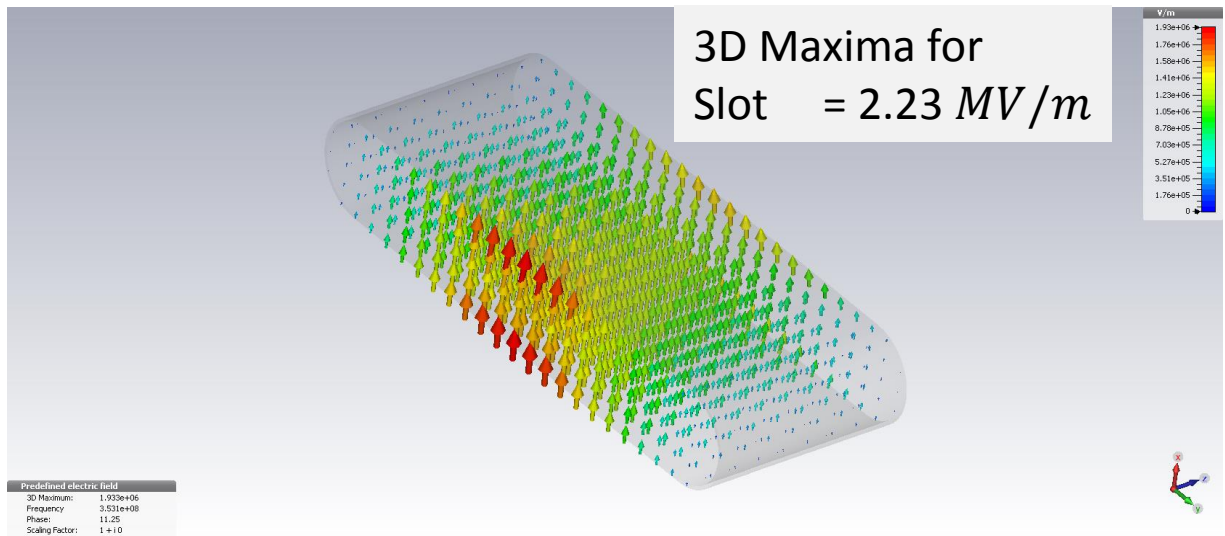

**Figure 8: Electric Field in Iris of TaCo**

*4.4.3.4  Second Order Multipacting Barrier*

The calculated barrier should occur around 0.0665 times the nominal field level, which corresponds to peak in <SEY> curve. This is further investigated with other post processing tools "particle vs time" and "emitted secondaries vs time" available in PIC (see Figure 9 and Figure 10).

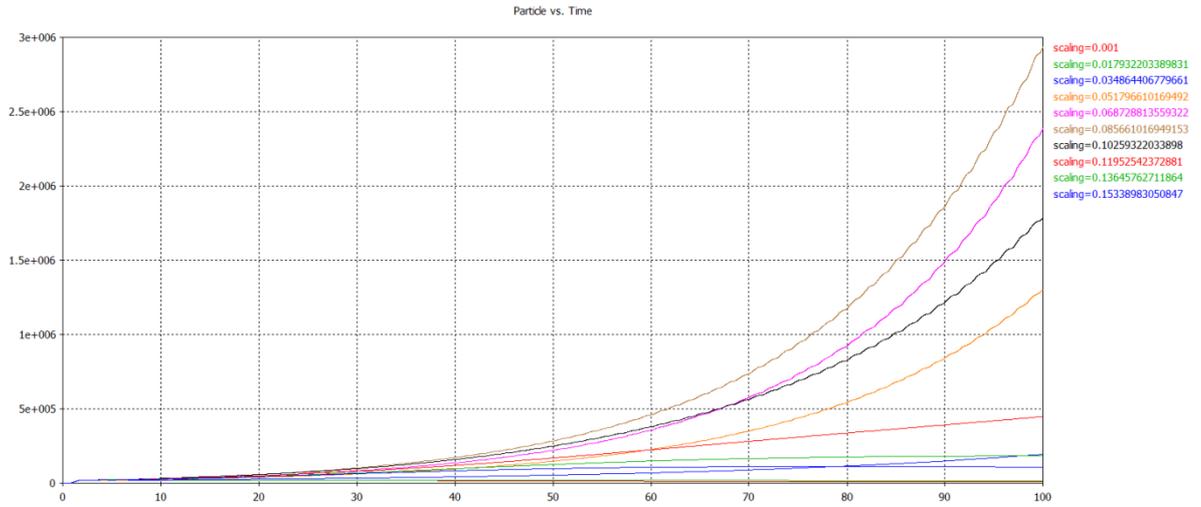

**Figure 9: Particle vs Time**

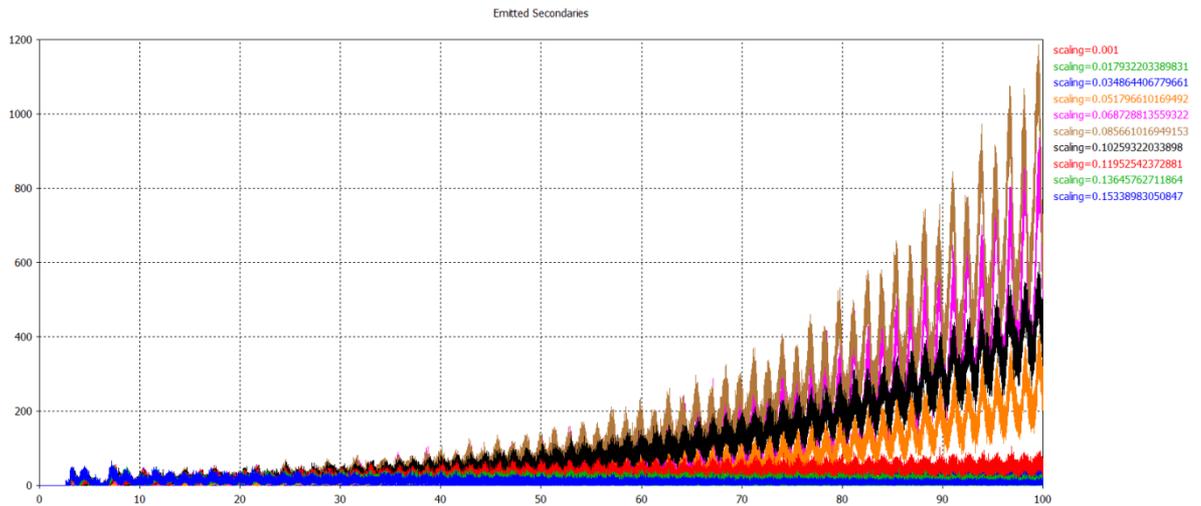

**Figure 10: Emitted Secondaries vs Time**

Careful observation of Figure 9 and Figure 10 clearly reveals that maximum generation of secondary particles takes place at 0.0856 times the maximum field in the iris and then it starts falling so that the 2$^{nd}$ order multipacting band is between 115.54E+3 V/m to 223.11E+3 V/m which corresponds to the field value calculated in Table 2. So the peak in Figure 7 corresponds to the two point 2$^{nd}$ order multipacting barrier. Another important observation stems from the fact that though particles are increasing with time the multiplication is not explosive.

*4.4.3.5    First Order Multipacting Barrier*

As calculated in Table 2, first-order-multipacting barrier should occur around 0.1996 times the maxima of iris field. But careful observation of <SEY> curve reveals no such peak around this field value. This is further investigated by using the post processing tools available in PIC (see Figure 11 and Figure 12).

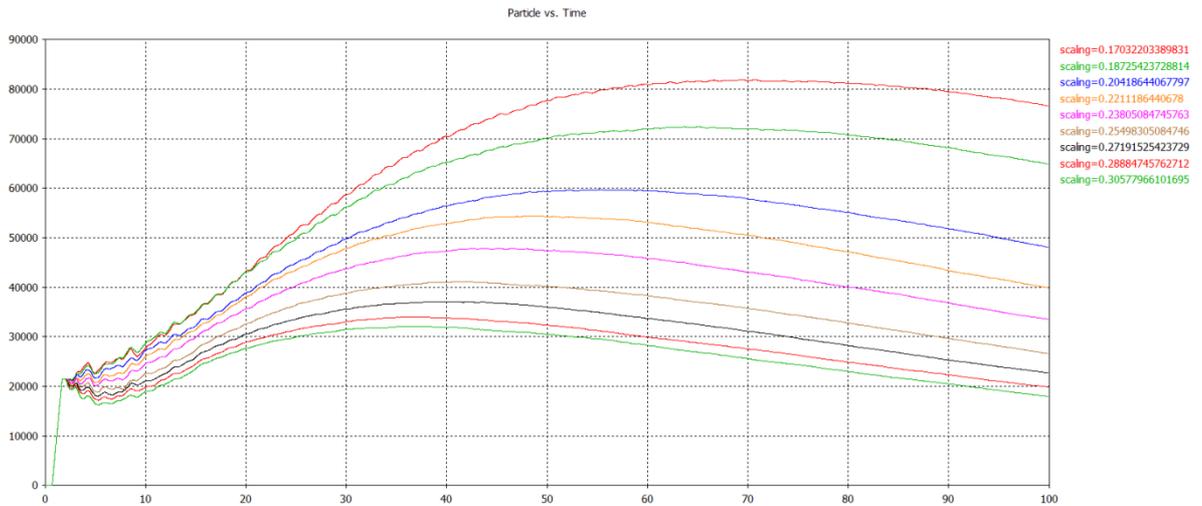

Figure 11: Particle vs Time

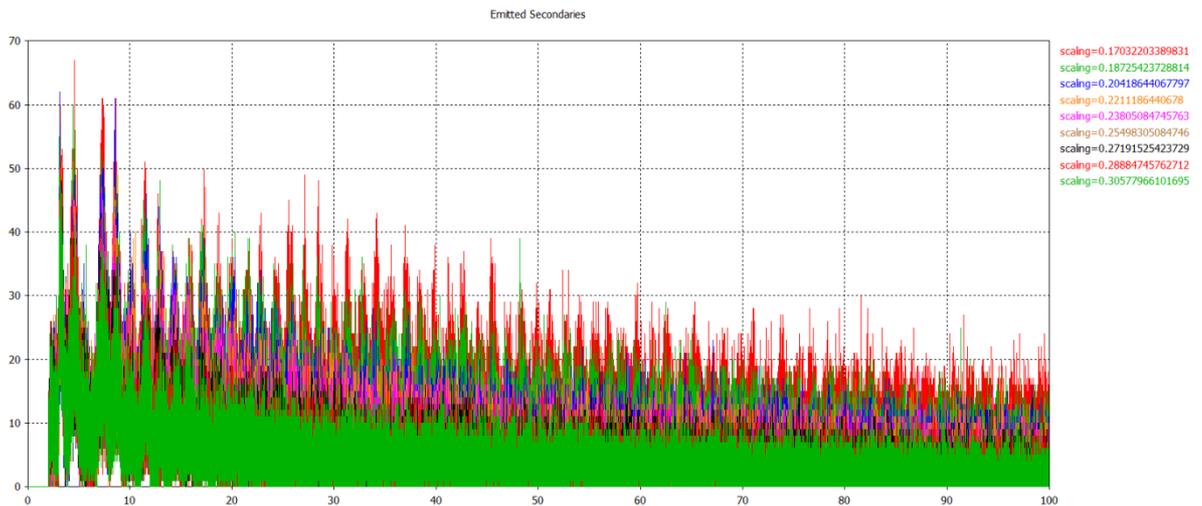

Figure 12: Emitted Secondaries vs Time

Careful observation of Figure 11 and Figure 12 confirms the absence of any resonance around the calculated first-order-multipacting barrier field value. Though in the start particles tend to increase and though there are significant collisions very soon the "particle vs time curve" shows a decreasing trend indicating the absence of any resonance.

4.4.3.5.1   Why is there no First Order Multipacting Barrier

What happens to particles at the field levels where we expect first-order-multipacting barrier can be explored with the "PIC position monitor" tool. It allows us not only to see the distribution of particles in the structure after a certain time but it also enables us to see their movement in the structure during the simulation.

From PIC position monitor tool it was very clear that as we approach the expected first-order-multipacting barrier it is not the electric field but the magnetic field which governs the multipacting process. After few oscillations particles are attracted towards the stronger magnetic field regions at the corner of the iris where they keep on oscillating. There the energy gain during the trajectory of the electrons is not sufficient to cause the emission of secondaries resulting in a reduction of particles and therefore we do not observe any multipacting. Figure 13 and Figure 14 shows the distribution of particles inside iris after 50 ns at two different field levels.

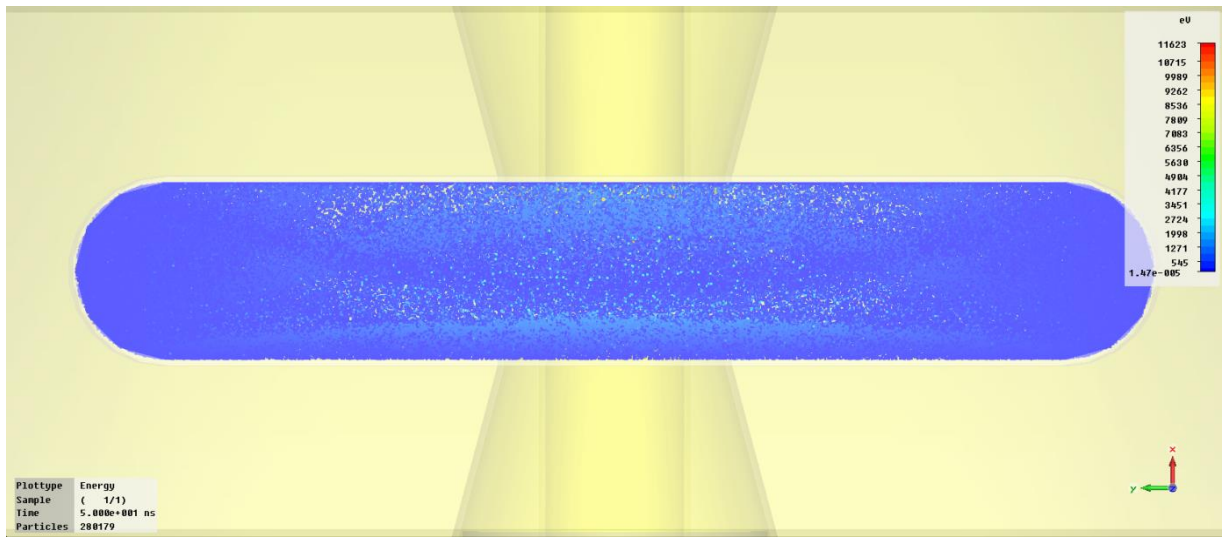

**Figure 13: Particle Distribution in Iris after 50ns @0.0939 times Field Maxima in Iris**

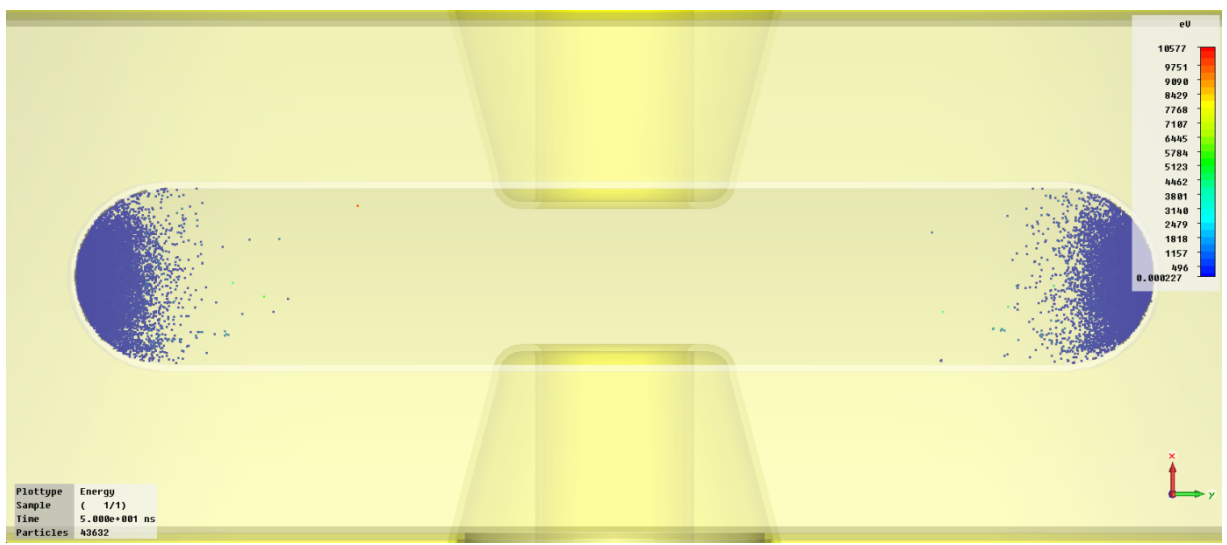

**Figure 14: Particle Distribution in Iris after 50ns @0.20 times Field Maxima in Iris**

*4.4.3.6 Around Operating Conditions*

<SEY> curve results were analysed near the operating field levels, which clearly indicates no danger of multipacting at operating conditions (see Figure 15 and Figure 16).

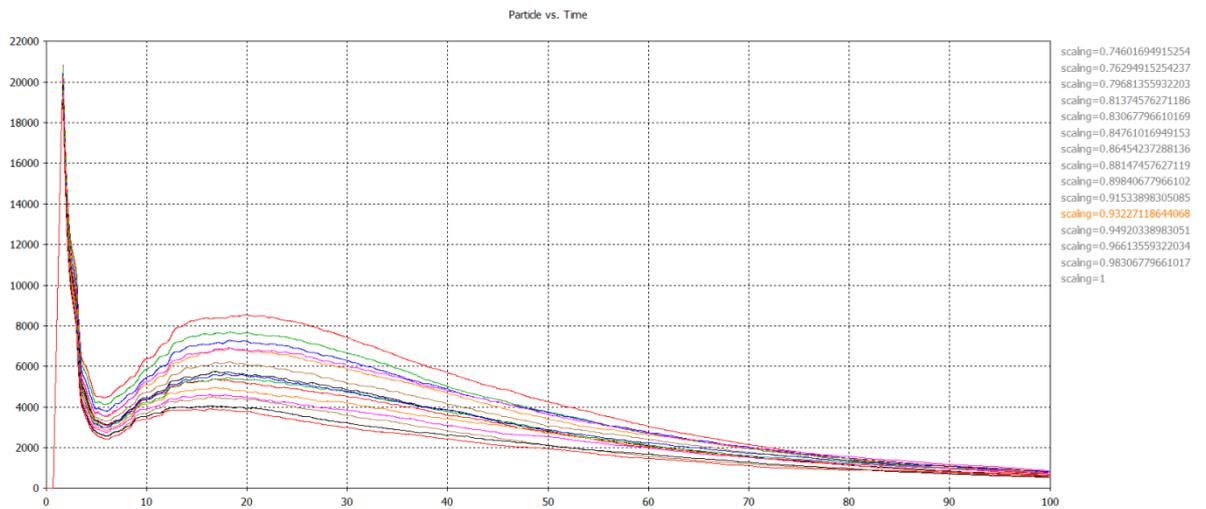

**Figure 15: Particle vs Time**

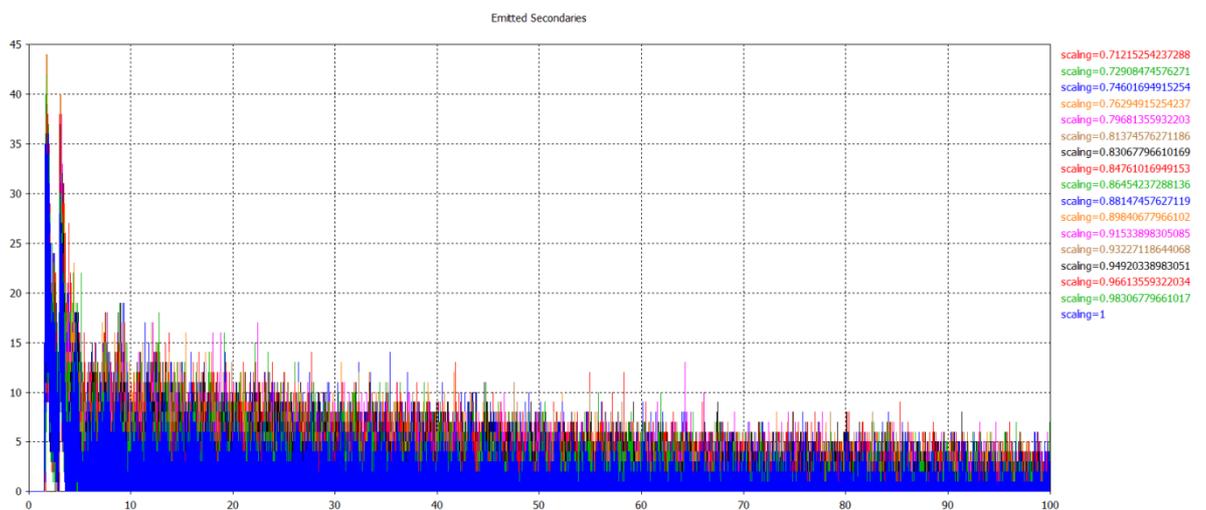

**Figure 16: Emitted Secondaries vs Time**

From Figure 15 and Figure 16 it is clear that there is no multipacting of any type around the operating conditions of the Tuner-adjustable Waveguide Coupler (TaCo), which is reflected in the <SEY> plot of Figure 7 showing <SEY> well below one meaning there is no multiplication of electrons. Figure 17 shows the distribution of particles inside iris of TACo after 50 ns at operating filed level.

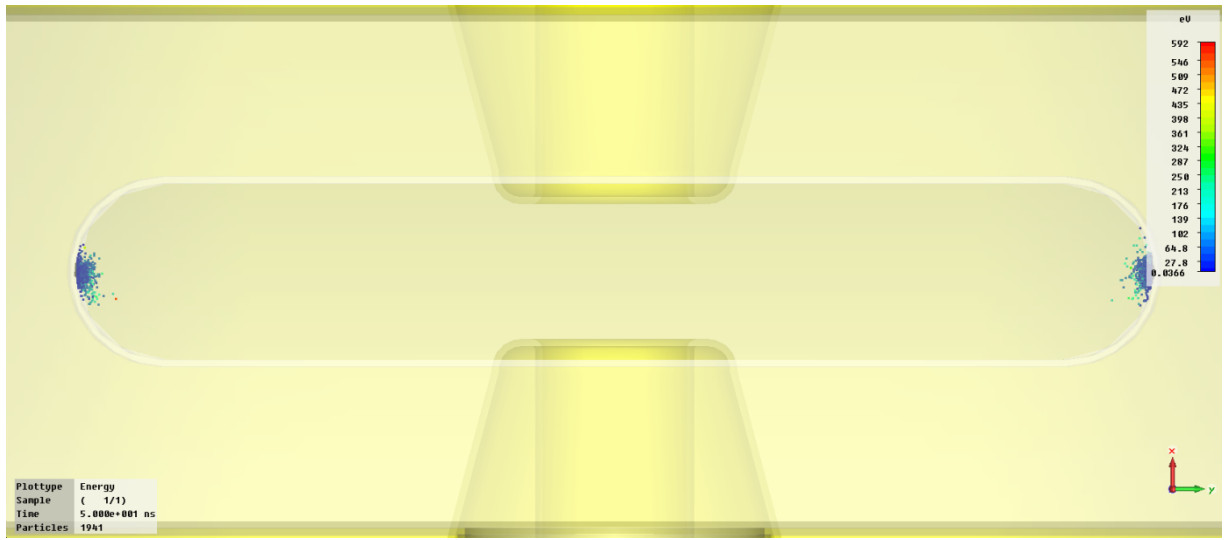

**Figure 17: Particle Distribution in Iris after 50ns @ Maximum Field in Iris**

## 5   Summary

A complete multipacting profile of the iris of the Linac4 Tuner-adjustable waveguide coupler (TaCo) was made with Particle Studio of CST using PIC Solver. Only two point second order multipacting barrier is encountered at a field range of 115.54E+3 V/m to 223.11E+3 V/m. The two-point first order multipacting barrier is supressed by the strong magnetic fields which govern the multipacting behaviour at that field value. Operating field values are safe for TaCo with no dangerous multipacting barrier occurring at or in the vicinity of operation.